\documentclass[preprint,12pt]{elsarticle}
\usepackage{stmaryrd}
\SetSymbolFont{stmry}{bold}{U}{stmry}{m}{n} 
\usepackage{amsfonts}
\usepackage{amsmath}
\usepackage{mathrsfs}
\usepackage{amssymb}
\usepackage{makeidx}
\usepackage{multirow}
\usepackage{algorithmic} 
\usepackage{bm}
\usepackage{setspace}
\usepackage{appendix}
\usepackage{amsthm}
\usepackage{empheq}
\usepackage{booktabs}
\usepackage[ruled,linesnumbered,vlined]{algorithm2e}  

    \SetCommentSty{mycommfont}
\usepackage{color}
\usepackage{enumerate}
\usepackage{makecell}
\usepackage{threeparttable}
\usepackage{booktabs}
\usepackage{tabularx}
\usepackage{array}
\usepackage[font=footnotesize]{caption}
\usepackage{tabularx,booktabs}
\usepackage[percent]{overpic}
\usepackage{setspace}

\usepackage{array} 
\usepackage{tikz,pgfplots}
\usetikzlibrary{spy} %
\pgfplotsset{width=8.6cm, height=6.5cm, compat=1.9}
 
\usepackage{url}
\usepackage{hyperref}

\journal{Computer Networks}

\begin{document}

\title{Mobility-Aware Decentralized Federated Learning with Joint Optimization of Local Iteration and Leader Selection for Vehicular Networks}

\author[inst1]{Dongyu Chen}
\author[inst1]{Tao Deng}
\author[inst1]{Juncheng Jia}
\author[inst1]{Siwei Feng}
\author[inst2]{Di Yuan}

\affiliation[inst1]{organization={School of Computer Science and Technology, Soochow University},
            country={China}}
\affiliation[inst2]{organization={Department of Information Technology, Uppsala University},
            country={Sweden}}

\begin{abstract}
Federated learning (FL) emerges as a promising approach to empower vehicular networks, composed by intelligent connected vehicles equipped with advanced sensing, computing, and communication capabilities.
While previous studies have explored the application of FL in vehicular networks, they have largely overlooked the intricate challenges arising from the mobility of vehicles and resource constraints. 
In this paper, we propose a framework of mobility-aware decentralized federated learning (MDFL) for vehicular networks. 
In this framework, nearby vehicles train an FL model collaboratively, yet in a decentralized manner.
We formulate a local iteration and leader selection joint optimization problem (LSOP) to improve the training efficiency of MDFL. 
For problem solving, we first reformulate LSOP as a decentralized partially observable Markov decision process (Dec-POMDP), and then develop an effective optimization algorithm based on multi-agent proximal policy optimization (MAPPO) to solve Dec-POMDP.
Finally, we verify the performance of the proposed algorithm by comparing it with other algorithms.
\end{abstract}


\maketitle



\section{Introduction}\label{sec:introduction}

With the rapid development of intelligent vehicles (IVs), an intelligent transportation system (ITS) greatly improves travel experience and safety. IVs surpass the scope of traditional transportation and gradually evolve into intelligent mobile platforms that integrate travel, information exchange, and decision assistance \cite{yan2024edge}. 
Nowadays, sensors installed on IVs, including LiDAR, RGB cameras, millimeter-wave radar, etc., will be able to capture massive environmental data. 
Effective utilization of these data can not only promote the deep application of machine learning in areas such as autonomous driving and intelligent traffic management, but also significantly impact the innovation and development of related algorithms, making ITSs more intelligent and efficient.

Constrained by the imperative of data privacy, a phenomenon of ``data islands" has gradually emerged among vehicles in vehicular networks.
This phenomenon restricts the effective integration and unified training of data.
To address the challenge, federated learning (FL), as a novel distributed machine learning paradigm, shows great potential in vehicular networks \cite{quemeneur2024fedpylot,Qu2024fedsa,Danish2024fedl}. 
With FL, vehicles achieve collaborative training of local models by exchanging only model parameters instead of directly transmitting user data. 
This eventually converges to a global model shared by all vehicles, promoting knowledge sharing and model optimization while ensuring data privacy\cite{posner2021federated}. 
The motivation for deploying FL in vehicular networks is as follows.
First, FL holds vast potential across a range of vehicular network scenarios, including traffic flow prediction (TFP), free-space detection (FSD), and driving behavior monitoring (DBM).
These applications significantly enhance the driving experience \cite{zhang2023federated}.
Secondly, FL deployment in vehicular networks safeguards privacy. The data gathered by vehicles often contain sensitive information, and FL ensures that such private data remain securely stored on the vehicles' onboard devices \cite{McMahan2017}.
Finally,  while vehicular mobility can pose communication challenges, the rapid movement of vehicles facilitates the blending of diverse data, which in turn boosts learning performance \cite{chen2024mobilityacc,shah2003data}.
The mobility of vehicles introduces dynamics in the client population during the model training phase. 
However, it potentially disrupts the successful transmission of model parameters among vehicles.
Apart from vehicle mobility, other challenges encompass bandwidth limitations, energy consumption, and transmission delay. 
However, most of existing works do not comprehensively consider the potential impact of vehicle mobility on FL performance, as well as the constraints imposed by limited resources, thus motivating us to address these issues. 

We investigate how to make the best use of vehicle mobility to design approaches for optimizing the training of FL models in vehicular networks, as well as how to enhance resource utilization efficiency.
Our objective is to maximize the accuracy of FL model.
The main contributions of this paper are as follows:

\begin{enumerate}
    \item We design a mobility-aware decentralized federated learning (MDFL) framework for vehicular networks. 
    Vehicles that are close to each other adopt a direct communication scheme to transmit data, while those farther apart utilize an indirect communication scheme to ensure communication stability. 
    To improve the training efficiency of MDFL, we formulate a local iteration and leader selection joint optimization problem (LSOP), taking into account the impact of vehicle mobility and resource constraints.   
    \item 
    For problem solving, we reformulate LSOP as a decentralized partially observable Markov decision process (Dec-POMDP).
    Subsequently, we develop an effective algorithm based on multi-agent proximal policy optimization (MAPPO) to solve Dec-POMDP.
    The algorithm involves two types of agents, that train their policy and value networks using centralized training and decentralized execution respectively, thereby enhancing decision-making quality.
    \item In performance evaluations, we utilize the simulation SUMO (simulation of urban mobility) to generate traffic flow data and train a LeNet network with the FashionMNIST data set. 
    We verify the effectiveness of the proposed algorithm by comparing it with other algorithms under various setting of parameters, including initial vehicle energy, reward proportional coefficients, indirect transmission energy consumption, and the number of vehicles. 
    In addition, we compare the performance of four classic aggregation strategies, i.e., FedAvg, FedNova, FedProx, and Scaffold, for both independent and identically distributed (IID) and Non-IID data distributions.  
\end{enumerate}

\section{Related work}

For vehicular networks, existing works on FL can be divided into two categories: centralized federated learning (CFL) and decentralized federated learning (DFL). 

\subsection{Centralized Federated Learning}

CFL, a classical federated learning paradigm, relies on a central server to aggregate the local model of clients. 
This requires additional physical nodes to serve as aggregation centers. 
Some works consider employing road side units (RSUs) as leaders \cite{Xie2022MOBFL,Zhang2023PALORA,Pervej2023Resource,Singh2024DRL-Based,chen2024mobilityacc,Xiang2024Collaborative,Macedo2023Multiple}.
The work in \cite{Xie2022MOBFL} assumes that the vehicle arrival process follows a Poisson distribution, and designs an effective method to optimize the duration of each training round and the number of local iterations. 
The work in \cite{Zhang2023PALORA} models FL as a joint scheduling problem of device selection, local round selection, and radio resource allocation.
The work in \cite{Pervej2023Resource} takes into account the residence time of vehicles near the edge server and include regularization terms in the aggregation strategy to make the contribution of vehicles with a longer residence time more significant to global model updates. 
The work in \cite{Singh2024DRL-Based} designs a deep reinforcement learning (DRL) caching decision model in which the vehicles with long stay times and low communication delays are assigned a higher aggregate weight. 
The work in \cite{chen2024mobilityacc} proposes a framework of hierarchical aggregation in which the mobility of vehicles is regarded as a shuffle of edge server datasets and the cloud server as aggregation nodes. 
With this framework, FL can maintain convergence even in a highly dynamic environment.
The work in \cite{Xiang2024Collaborative} models vehicle mobility as the probability of model transmission failure and optimizes local training rounds to address the risk of communication disruption. 
For multi-RSU networks, the work in \cite{Macedo2023Multiple} proposes a framework
where vehicles will alert the RSU for update when significant model deviations are detected, and vehicles that are about to move out of the communication range are ignored.
The works in \cite{Xie2022MOBFL,Zhang2023PALORA,Pervej2023Resource,Singh2024DRL-Based,chen2024mobilityacc,Xiang2024Collaborative,Macedo2023Multiple} rely on RSUs to alleviate the issues of frequent data domain switching and communication resource shortages caused by high vehicle mobility. 
However, this solution heavily relies on the hardware and software implementation of the RSU, which inevitably limits the scope of FL training to the RSU's coverage area. Moreover, the availability of an RSU is not guaranteed, and it becomes a single point of failure, and thus a malfunction, maintenance, or insufficient availability can result in a complete outage of the entire system's service within the region \cite{Savazzi2020DFLIOT}.

\subsection{Decentralized Federated Learning}

In contrast to CFL, DFL does not rely on a fixed central server, but flexibly assigns the task of model aggregation to clients. 
According to the distribution of aggregation tasks, DFL can be divided into fully DFL (FDFL) 
\cite{liu2022enhance} and leader-follower DFL 
\cite{AbdulRahman2023,Tan2024SC1BCS,Yang2022LeadFN,behera2021federated,Gou2024Vote} built on peer-to-peer networks.
In FDFL, each client is not only involved in the training of the local model, but also responsible for collecting and aggregating models from its neighbor clients. 
The work in \cite{liu2022enhance} assumes that vehicles directly communicate with their nearest RSU, and the global model is aggregated at the RSU level. 
This design significantly reduces the impact of vehicle movement on the model training process. 
Although the above work have proposed some effective methods to improve the efficiency of model training, they face the challenge of high communication demand. 

In the leader-follower DFL, a client from a client group is dynamically selected to act as the executor of the aggregation task (i.e., the leader). 
This design retains a server-like logic, thus it can alleviate spatial constraints and significantly reduce the communication pressure of FDFL. 
The leader's selection strategy has a decisive impact on system efficiency of the leader-follower DFL.
Some works propose effective algorithms to optimize the selection of the leader \cite{AbdulRahman2023,Tan2024SC1BCS,Yang2022LeadFN,behera2021federated,Gou2024Vote}.
The work in \cite{AbdulRahman2023} proposes a vehicle clustering strategy based on a clustering algorithm.
The strategy evaluates the vehicles in the cluster using a quality of service (QoS) scoring mechanism and cleverly distributes the communication load of RSU to the vehicles among the cluster heads, effectively alleviating the communication pressure on the RSU. 
In \cite{Tan2024SC1BCS}, the vehicles in each cluster transmit models compressed by adaptive thresholds to the cluster head for aggregation. 
The work in \cite{Yang2022LeadFN} identifies that selecting a device with high computing, communication, and energy supply capabilities as the leader can effectively manage model aggregation and accelerate the FL process. 
The work in \cite{behera2021federated} designs a leader election algorithm based on fair voting among devices using the raft consensus mechanism. 
The work in \cite{Gou2024Vote} 
quantifies the probability of successfully electing a leader, the efficiency of communication, and the accuracy of joint decision-making through mathematical derivation.
However, these works lack comprehensive consideration of the dynamic changes in the real-time position of the vehicles, as well as resource constraints. 

\section{System model and problem formulation}

\begin{figure*}[!t]
  \centering
  \includegraphics[width=1\textwidth]{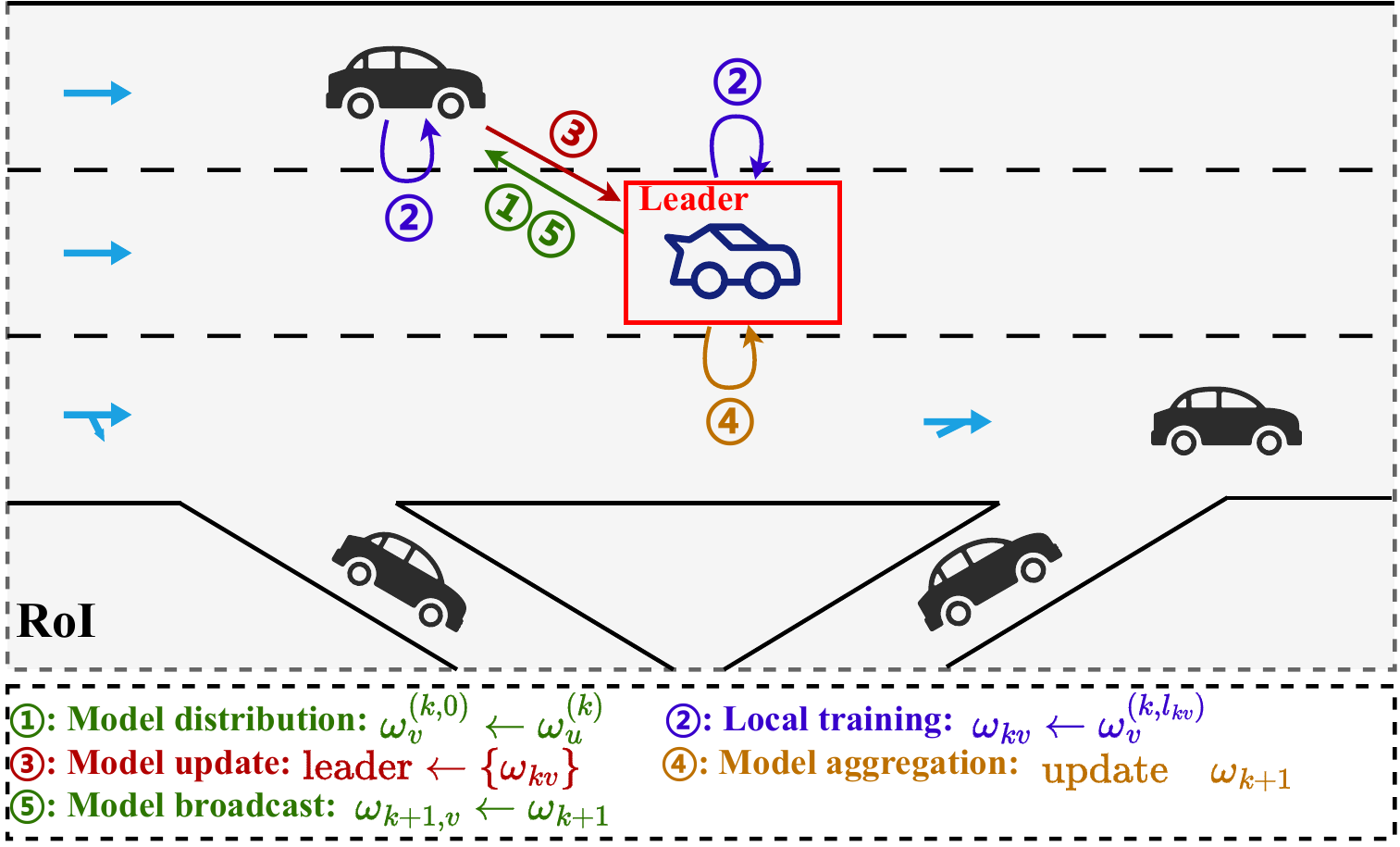}\\
  \caption{MDFL framework.}
  \label{fig:system}
\end{figure*}
\subsection{MDFL framework}
Fig. \ref{fig:system} shows the MDFL framework that describes a two-dimensional rectangular region of interest (RoI).
In the RoI, vehicles are assumed to have reached a consensus and jointly train a global MDFL model, denoted by $\boldsymbol{\omega}$ \cite{liu2023reputation}.
Vehicles outside of this region exit the model training process.
Denote by $\mathcal{N}$ the set of vehicles in the MDFL framework.
Denote by $K$ the total number of communication rounds of MDFL.
In the $k$-th communication round, $k\le K$,
the set of vehicles in the RoI is denoted by $\mathcal{N}_k$. 
Vehicle $v$, $v\in\mathcal{N}_k$, shares information with other vehicles in the RoI, including coordinates, speed, and acceleration, denoted by $(x_v,y_v)$, $s_v$, and $a_v$, respectively.
In $(x_v,y_v)$, $x_v$ and $y_v$ represent the longitude and latitude coordinates of vehicle $v$, respectively.
Due to vehicle mobility, the set of available vehicles $\mathcal{N}_k$ varies by $k$, such as vehicles being in the RoI and having energy for model training in the $k$-th communication round.
Denote by $t^{round}$ the time duration of a communication round.
We assume that all the vehicles in the RoI complete their local model training and transmission within this time duration.
Each vehicle dynamically adjusts its local training rounds based on its distance from the selected server node (called the leader), to ensure the effectiveness of the overall training process, thereby adapting to the dynamic changes in the vehicle distribution and communication conditions within the area.

\subsection{MDFL Training Model}
The model training process of MDFL employs a leader-follower strategy. 
In each communication round, one vehicle in the RoI, designated at the leader, as shown in Fig.\ref{fig:system}, is responsible for model dissemination and aggregation,
and the remaining vehicles are called followers.
In this paper, we use the terms vehicle and follower interchangeably.
All the vehicles, including the leader, utilize their local data for local training. 
The local data of a vehicle is the data collected by it before entering the RoI (for example, in an autonomous driving task, using road image information to predict the rotation angle of the steering wheel).
The training process consists of the following five parts.

\begin{enumerate}
    \item \textbf{Model distribution}: At the beginning of the $k$-th communication round, the leader $u$ sends its local model, denoted by $\omega_u^k$, to the followers.
    Thus, the initial local model of any follower $v$ in the $k$-th communication round, denoted by $\omega_v^{(k,0)}$, is expressed as
        \begin{equation}
    \begin{aligned}
    \omega_v^{(k,0)}=\omega_u^k, ~~v\in \mathcal{N}_k \setminus u. \\
    \end{aligned}
    \label{eq:Initial Model}
    \end{equation}
    \item \textbf{Local training}: Denote by ${l}_{kv}$ the number of local iterations of follower $v$ in the $k$-th communication round.
    After ${l}_{kv}$ local iterations, the model of follower $v$, denoted by $\omega_v^{(k,{l}_{kv})}$,  is expressed as
    \begin{equation}
    \begin{aligned}
    \omega_v^{(k,{l}_{kv})}=\omega_v^{(k,0)}-\eta\sum_{l=1}^{{l}_{kv}}{\nabla}f_v(\omega_v^{(k,l)}),
    \end{aligned}
    \label{eq:loss-define}
    \end{equation}
    where $\eta$ represents the learning rate, and ${\nabla}f_v(\omega_v^{(k,l)})$ represents the gradient of the local model loss function.
    \item \textbf{Model update}: 
    The followers in the RoI send their local models $\omega_v^{(k,l_{kv})}$ to the leader.
    \item \textbf{Model aggregation}: The leader aggregates all the received models by using a weighted average method,
    \begin{equation}
    \begin{aligned}
    \omega_u^{(k+1)}=\sum_{v\in\mathcal{N}_k}\beta_v\omega_v^{(k, {l}_{kv})}.
    \end{aligned}
    \label{eq:aggregation-define}
    \end{equation}
    where $\beta_v$ represents the aggregate weight, which satisfies $\sum_{v\in\mathcal{N}_k}\beta_v=1$. The specific aggregate weight allocation depends on different aggregation strategies.
    \item \textbf{Model broadcast}: The leader resends the aggregated model $\omega_u^{(k+1)}$ to the followers in the RoI.
    The training process repeats until the loss function converges or the communication round reaches the upper bound.
\end{enumerate}

\begin{figure*}[!t]
  \centering
  \includegraphics[width=1.05\textwidth]{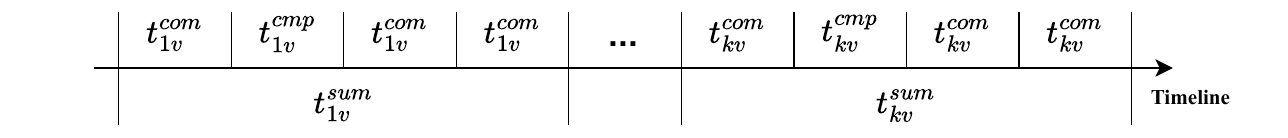}\\
  \caption{Timeline of MDFL.}
  \label{fig:timeline}
\end{figure*}

Fig. \ref{fig:timeline} describes the timeline of follower $v$ during the training process of MDFL. 
We assume that the vehicular network divides time into fixed intervals, with all operations and communications occurring within these intervals \cite{yan2024dynamic}.
In the $k$-th communication round, denote by $t_{kv}^{com}$ the total time of model distribution, model update, and model broadcast processes.
Denote by $t^{cmp}_{kv}$ the time spent on local training. 
Denote by $t_{kv}^{sum}$ the total time of a communication round. 
The aggregation process is not incorporated into our timeline illustration due to its comparatively brief duration.


\subsection{Resource Consumption Model}
\begin{figure}[!t]
    \centering
    \includegraphics[width=3in]{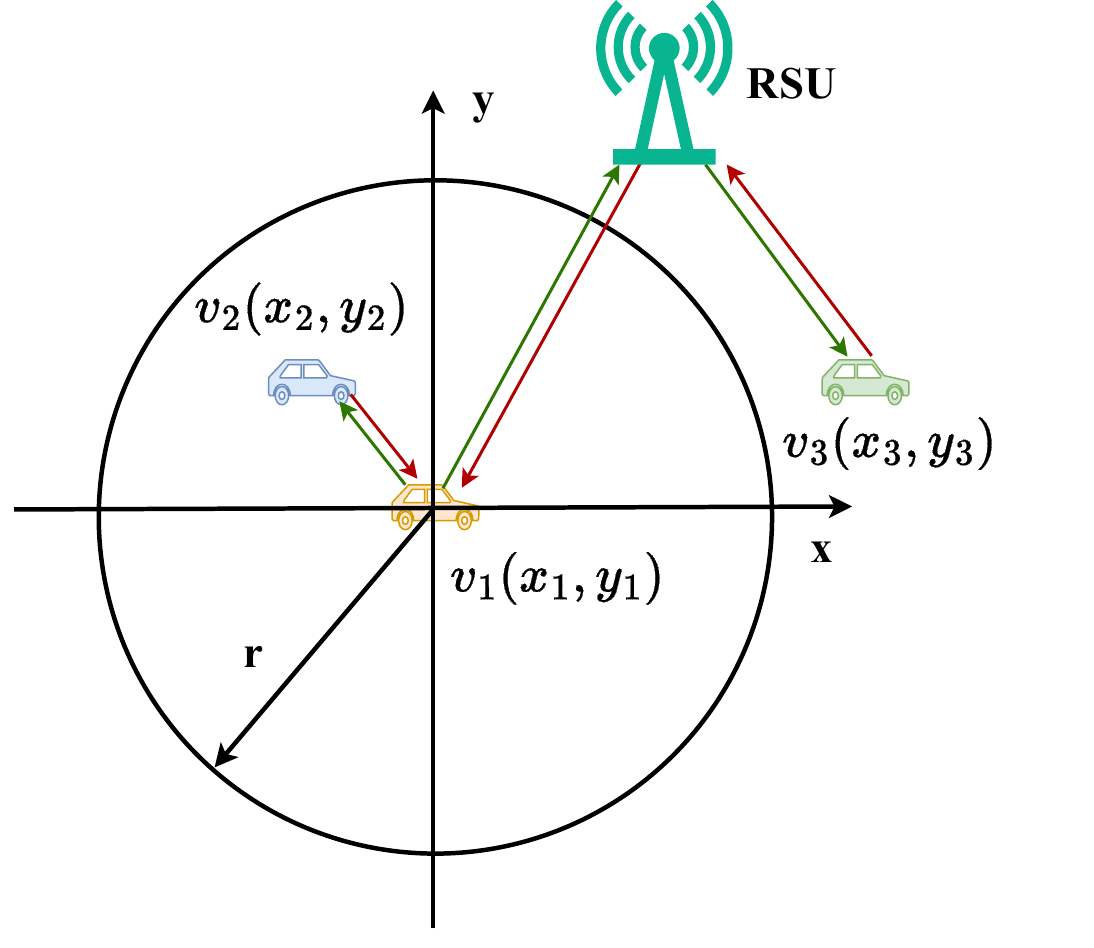}
    \caption{Communication model of vehicles.}
    \label{fig:region}
\end{figure}

The communication model between vehicles is shown in Fig. \ref{fig:region}. 
The direct communication range of any two vehicles is defined as a circular area with a radius of $r$. 
In order to ensure that the model parameters are transmitted to the specified vehicle timely, vehicles outside this range have to rely on cellular networks to transmit data to each other.
This process is referred to as indirect communication transmission. 
We use Euclidean distance to define the relative distance between vehicles. 
The distance between vehicle $i$ and vehicle $j$, denoted by $d_{ij}$, is expressed as
\begin{equation}
\begin{aligned}
d_{ij} = \sqrt{|x_i - x_j|^2 + |y_i - y_j|^2}.\\
\end{aligned}
\label{eq:euclidean-distance}
\end{equation}
Denote by $p_{kij}^{\mathrm{suc}}$ the probability of using direct communication between vehicles $i$ and $j$ in the $k$-th communication round, which is expressed as
\begin{equation}
\begin{aligned}
p_{kij}^{\mathrm{suc}}=\mathcal{P}(d_{kij} \le r).\\
\end{aligned}
\label{eq:probability-actual}
\end{equation}
Denote by $E_{edge}^C$ and $T^C_{edge}$ the energy and time consumed by direct communication transmission, respectively.
Denote by $E^C_{cloud}$ and $T^C_{cloud}$ the energy and time consumed by indirect communication transmission, respectively. 
Thus, in the $k$-th communication round, 
for the leader $u$, the energy consumption is expressed as
\begin{equation}
\begin{aligned}
e_{ku}^{com}=
\sum_{v \in {\mathcal{N}_k} \setminus u} \left [ p_{kuv}^{\mathrm{suc}}E_{edge}^C+(1-p_{kuv}^{\mathrm{suc}})E_{cloud}^C \right ] .\\
\end{aligned}
\label{eq:energy-actual-leader}
\end{equation}
For a follower $v$, $v \in {\mathcal{N}_k} \setminus u$, the energy consumption is expressed as
\begin{equation}
\begin{aligned}
e_{kv}^{com}=p_{kuv}^{\mathrm{suc}}E_{edge}^C+(1-p_{kuv}^{\mathrm{suc}})E_{cloud}^C.\\
\end{aligned}
\label{eq:energy-actual-follower}
\end{equation}
The actual time required by the leader $u$ is expressed as
\begin{equation}
\begin{aligned}
t_{ku}^{com}=
\max\limits_{v \in \mathcal{N}_k \setminus u} \left [ p_{kuv}^{suc}T^C_{edge}+(1-p_{kuv}^{suc})T^C_{cloud} \right ].\\
\end{aligned}
\label{eq:time-actual-leader}
\end{equation}
The actual time required by the follower $v$ is expressed as
\begin{equation}
\begin{aligned}
t_{kv}^{com}=
p_{kuv}^{suc}T^C_{edge}+(1-p_{kuv}^{suc})T^C_{cloud}.\\
\end{aligned}
\label{eq:time-actual-follower}
\end{equation}

In a single communication round, vehicle $v$ needs to go through three communication processes: model distribution, model update, and model broadcast processes, each of which consumes energy $e^{com}_{kv}$.
In addition, the vehicle also needs to undergo a local training process, which consumes energy $e^{cmp}_{kv}$.
Denote by $e^{sum}_{kv}$ the total energy consumption of vehicle $v$ in the $k$-th communication round, which is expressed as
\begin{equation}
\begin{aligned}
e^{sum}_{kv} &= e^{cmp}_{kv} + 3e^{com}_{kv}.
\end{aligned}
\label{eq:energy-total}
\end{equation}
In (\ref{eq:energy-total}), 
\begin{equation}
\begin{aligned}
 e^{cmp}_{kv} = e^{cmp}_{itr}{l}_{kv},
\end{aligned}
\label{eq:energy-total-train}
\end{equation}
where $e^{cmp}_{itr}$ represents the energy consumption per local iteration. 
Similar to energy consumption, the total time consumption $t^{sum}_{kv}$ is expressed as
\begin{equation}
\begin{aligned}
t^{sum}_{kv} &= t^{cmp}_{kv} + 3t^{com}_{kv}.
\end{aligned}
\label{eq:time-total}
\end{equation}
In (\ref{eq:time-total}), 
\begin{equation}
\begin{aligned}
t^{cmp}_{kv} = t^{cmp}_{itr}{l}_{kv},
\end{aligned}
\label{eq:time-total-train}
\end{equation}
where $t^{cmp}_{itr}$ represents the time duration per local training. 

In (\ref{eq:energy-actual-leader}) and (\ref{eq:energy-actual-follower}),  $e^{com}_{kv}$ is used to calculate the actual communication energy consumption of the vehicle at the end of a communication round. 
As the number of local iterations and leader selection are jointly decided at the beginning of a communication round, it is important to predict the vehicle's relative position after completing its local iterations. 
The horizontal axis is the main direction of mobility in our system scenario.
Thus, we utilize the training time $t_{k-1,v}^{cmp}$ from the previous communication round of follower $v$, and its x-coordinate $x_v$, speed $s_v$ and acceleration $a_v$.
We also consider the x-coordinate $x_u$, speed $s_u$ and acceleration $a_u$ of the leader $u$ to make a worst-case estimate of the distance $d'$ travelled by the vehicles.

Specifically, at the beginning of the $k$-th communication round, each vehicle predicts its position.
By leveraging this predicted position, the vehicle then computes its estimated energy consumption $e'$ in the $k$-th communication round.
This predictive approach enables vehicles to more effectively modulate the quantity of local iterations.
Denote by $d'_{kuv}$ the predicted relative position between vehicles $u$ and $v$, which is expressed as
\begin{equation}
\begin{aligned}
d'_{kuv}=d_{kuv}+(x_u-x_v)({\Delta}d_u-{\Delta}d_v),\\
\end{aligned}
\label{eq:distance-expect}
\end{equation}
where 
\begin{equation}
\left\{
\begin{aligned}
{}&\Delta d_u=s_ut^{cmp}_{k-1,v}+\frac{1}{2}a_u({t^{cmp}_{k-1,v}})^2,\\
{}&{\Delta}d_v=s_vt^{cmp}_{k-1,v}+\frac{1}{2}a_v({t^{cmp}_{k-1,v}})^2.
\end{aligned}
\right.
\label{eq:distance-expect11}
\end{equation}
For the first communication round, $t_{k-1,v}^{cmp}$ is 0. 
Denote by ${p'}_{kuv}^{\mathrm{suc}}$ the predicted probability of direct communication between vehicles $u$ and $v$ in the $k$-th round, which is expressed as
\begin{equation}
\begin{aligned}
{p'}_{kuv}^{\mathrm{suc}}=\mathcal{P}(d'_{kuv} \le r).\\
\end{aligned}
\label{eq:probability-expect}
\end{equation}
Denote by $e'^{com}_{ku}$ and $e'^{com}_{kv}$ the predicted communication energy consumption by the leader $u$ and a follower $v$ in the $k$-th round, respectively, which are expressed as
\begin{equation}
\left\{
\begin{aligned}
{}&{e'}_{ku}^{com}=
\sum_{v \in \mathcal{N}_k \setminus u} \left [ {p'}_{kuv}^{\mathrm{suc}}E_{edge}^C+(1-{p'}_{kuv}^{\mathrm{suc}})E_{cloud}^C \right ] ,\\
{}&{e'}_{kv}^{com}=
{p'}_{kuv}^{\mathrm{suc}}E_{edge}^C+(1-{p'}_{kuv}^{\mathrm{suc}})E_{cloud}^C.
\end{aligned}
\right.
\label{eq:distance-expect00}
\end{equation}

\subsection{Leader Selection Model}
\label{subsection:leader}
Denote by $R_{kv}$ a binary variable with $\sum_{v\in\mathcal{N}_k} R_{kv}=1$, $R_{kv} \in \{0,1\}$, 
which is one if and only if vehicle $v$ is selected as the leader in the $k$-th communication round, otherwise zero.
The selected vehicle $v$ is denoted by $v^*$.
Denote by $\boldsymbol{R}$ the corresponding matrix, where $\boldsymbol{R}_k$ is a vector containing the leader selection status of all vehicles $v\in\mathcal{N}_k$.
In order to select an optimal leader in each communication round, we design a leader election mechanism based on energy and participation.
The fundamental premise underlying this mechanism pertains to the leader's energy resources, which is pivotal in ensuring the seamless execution of both communication and computational tasks.
Therefore, to evaluate potential leader candidates, we use an energy ratio as the central index, which is directly related to the residual energy level of the vehicle. Denote by $E_v$ the initial energy of vehicle $v$. Denote by $W(\boldsymbol{R}_k)$ the energy ratio $W$ in the $k$-th communication round, which is expressed as
\begin{equation}
\begin{aligned}
W(\boldsymbol{R}_k)=\sum_{v\in\mathcal{N}_k}
\frac{e_{kv}^{res}}{\sum\limits_{v'\in\mathcal{N}_k}e_{kv'}^{res}}
{R}_{kv},\\
\end{aligned}
\label{eq:rate-energy}
\end{equation}
where $e_{kv}^{res}$ represents the residue energy of vehicle $v$ in the $k$-th communication round, which is expressed as
\begin{equation}
\begin{aligned}
e_{kv}^{res}=E_v-\sum_{j=1}^{k-1}e^{sum}_{jv}.\\
\end{aligned}
\label{eq:res-energy}
\end{equation}

In addition, the introduction of newly added vehicles can potentially disrupt the training process prior to the convergence of the global model, while all vehicles experience a gradual depletion of their energy reserves throughout the training progresses. Thus, during the vehicle's participation in model training, the trend of a vehicle being selected as the leader first increases and then decreases. 
In this paper, we propose a participation ratio adjustment strategy based on the Gaussian function.
Specifically, in the $k$-th communication round, the participation ratio $G$ is expressed as
\begin{equation}
\begin{aligned}
G(\boldsymbol{R}_k)=\sum_{v\in\mathcal{N}_k}
e^{-\frac{1}{2}(h_{kv}-\rho)^2}
{R}_{kv},\\
\end{aligned}
\label{eq:rate-participation}
\end{equation}
where $h_{kv}$ represents the number of communication rounds up to the $k$-th one, in which vehicle $v$ has participated in training, and $\rho$ represents a hyperparameter determined based on the actual running process. This strategy aims to dynamically adjust the participation ratio of each vehicle to address the challenges posed by energy depletion and new participants, to ensure the stability and efficiency of the training process.
MDFL ensures fairness and mitigates the communication burden on the leader in three aspects.
First, it employs a mechanism of model distribution, update, and broadcast. This not only prevents model updates from being overwritten but also synchronizes these updates across all vehicles, ensuring that each vehicle receives the updates equitably. 
Second, MDFL dynamically designates vehicle leaders based on a combination of energy levels and participation rates, thereby reducing the communication load on the leaders.
Finally, follower vehicles are encouraged to conduct as many local iterations as possible within each communication round, which enhances the efficiency of model updates.

\subsection{Problem formulation}
Our problem is to maximize the accuracy of the global model.
Denote by $\mathcal{F}(\boldsymbol{w}_v^{K},\mathcal{D}^{\mathrm{val}}_v)$ the accuracy of the local model of vehicle $v$, i.e., $\boldsymbol{w}_v^{K}$, on its validation set $\mathcal{D}^{\mathrm{val}}_v$ in the final communication round.
Denote by $F^{acc}$ the average of $\mathcal{F}(\boldsymbol{w}_v^{K},\mathcal{D}^{\mathrm{val}}_v)$ of vehicles within the RoI during the final communication round indexed by $K$, which is expressed as
\begin{equation}
\begin{aligned}
F^{acc}=\frac{1}{N_{K}}\sum_{v\in\mathcal{N}_{K}}\mathcal{F}(\boldsymbol{w}_v^{K},\mathcal{D}^{\mathrm{val}}_v).
\end{aligned}
\label{eq:final-accuracy}
\end{equation}
The local iteration and leader selection joint optimization problem (LSOP) is given in (\ref{eq:final-target}).
\begin{figure}[htbp]
\begin{subequations}
\begin{alignat}{2}
& \max\limits_{\boldsymbol{R},\boldsymbol{l}}F^{acc} \label{a} \\
\text{s.t}. \quad
& \begin{aligned}
    e^{cmp}_{kv}({l}_{kv}) + 3{e'}_{kv}^{com}({R}_{kv},{l}_{k-1,v}) &\\
    \le e_{kv}^{res}, \quad v \in \mathcal{N}_k, \, k \le {K},
    \end{aligned} \label{b} \\
& e_{kv}^{sum}({R}_{kv},{l}_{kv}) \le e_{kv}^{res}, v\in\mathcal{N}_k,k\le{K}, \label{c}\\
& t^{sum}_{kv}({R}_{kv},{l}_{kv})\le t^{round},v\in\mathcal{N}_k,k\le{K}, \label{d}\\
& R_{kv^*}= \arg \max \limits_{\boldsymbol{R}_k} \left [W(\boldsymbol{R}_k)+{\epsilon}G(\boldsymbol{R}_k) \right ],k\le{K}, \label{e}\\
& {l}_{kv}\in\mathbb{Z}^+,k\in\mathbb{Z}^+,K\in\mathbb{Z}^+ \label{f},N_k\ge2.
\end{alignat}
\label{eq:final-target}
\end{subequations}
\end{figure}

In (\ref{eq:final-target}), Constraints (\ref{b}) and (\ref{c}) ensure that the predicted and actual energy consumption of each participating vehicle in any communication round does not exceed its current energy capacity, respectively.
Constraint (\ref{d}) ensures that the time spent by each vehicle in any communication round is strictly bounded by a time duration of $t^{round}$. 
Constraint (\ref{e}) specifies the leader selection method in the MDFL framework, and $\epsilon$ represents the trade-off coefficient. 
Constraint (\ref{f}) address the validity of the parameters.

\section{Algorithm design}

Solving the problem in (\ref{eq:final-target}) faces three challenges.
First, the two sets of optimization variables $\boldsymbol{l}$ and $\boldsymbol{R}$ are strongly coupled with each other. The values of one set is not only constrained by its own parameters and constraints, but is also affected by the other set, and vice versa.
Second, the set of vehicles $\mathcal{N}_k$ in each round $k$ is time-varying, as vehicles may dynamically enter or exit the RoI at different time points. 
Traditional optimization methods have fixed model parameters and lack real-time adaptability, making them unable to adapt to the instantaneous changes in vehicle mobility. 
The limitation of these methods lies in their inability to fully capture the space-time dynamics of vehicle movement patterns, which restricts their application in complex vehicular network scenarios. 
Finally, the lack of a unified global scheduling mechanism results in significant heterogeneity and inconsistency of traffic information observed among vehicles.
This phenomenon not only increases the complexity of the vehicle networking system, but also requires each vehicle to make autonomous decisions as an independent agent in the absence of a global view. Therefore, each vehicle needs to rely on its local sensing capabilities, historical data, and inter-vehicle communication to construct the local information required for its decision-making.

To address these challenges, we propose an effective local iteration and leader selection algorithm based on multi-agent proximal policy optimization (MAPPO) \cite{Yu2021MAPPO}. 
The proposed algorithm integrates global information by introducing a centralized value function, enabling all independent agents to fully utilize this global perspective in the decision-making process.
Specifically, we first define two types of agents, then reformulate the LSOP problem as a decentralized partially observable Markov decision process (Dec-POMDP), and finally design a MAPPO-based algorithm to solve Dec-POMDP.

\subsection{Selector Classification}

\begin{figure}[!t]
    \centering
    \includegraphics[width=4in]{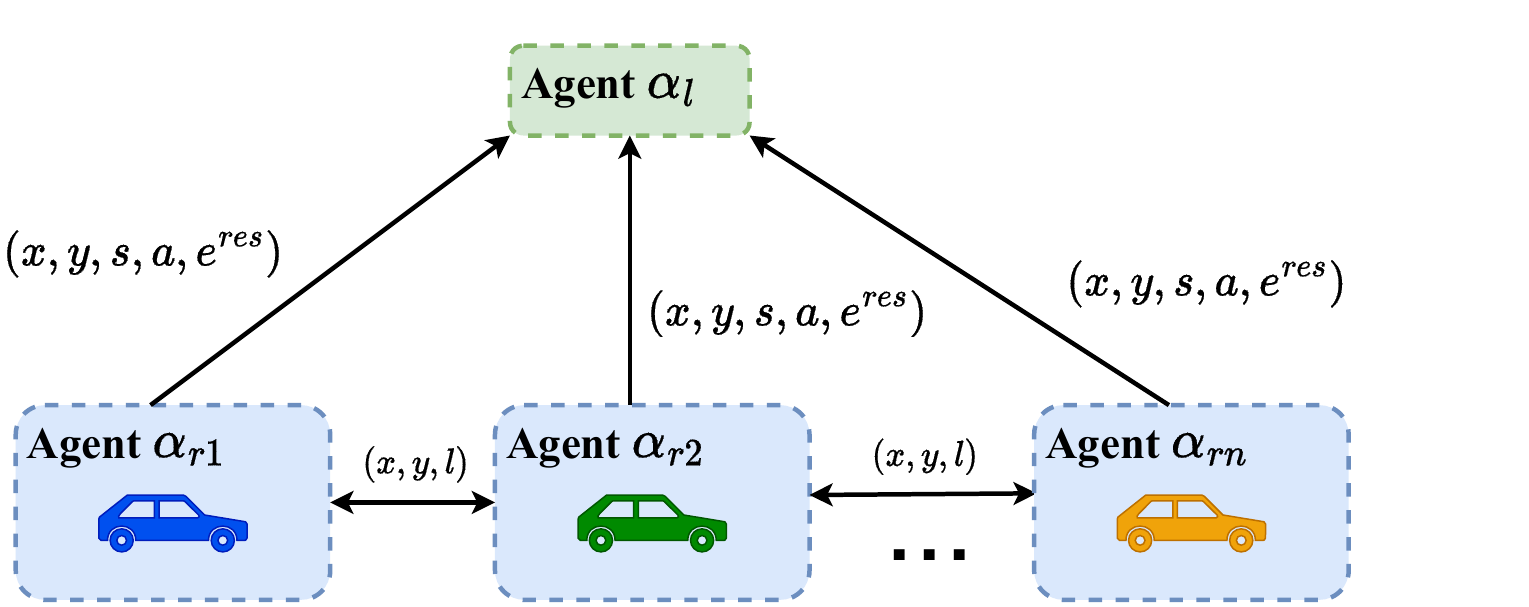}
    \caption{The relationship of agents.}
    \label{fig:agents}
\end{figure}

To solve LSOP with two sets of optimization variables, we design one type of selector for each variable set.
These selectors, as agents in the environment, make joint decisions for problem-solving.
Specifically, we design two types of agents, as shown in Fig. \ref{fig:agents}:



\begin{itemize}
    \item \textbf{Local iteration selector} $\alpha_r$: Each vehicle integrates a local iteration selector mechanism, responsible for monitoring the position information of other vehicles within the RoI and their respective local training iterations. With (\ref{eq:euclidean-distance}), the selector can accurately calculate the current position of the target vehicle, thereby constructing a local observation environment. 
    Based on this observation, the selector makes decisions for each vehicle's local training in every communication round.
    \item \textbf{Leader selector} $\alpha_l$: 
    The agent first comprehensively observes and collects key information such as the position, speed, and energy status of all vehicles. 
    It then utilizes (\ref{eq:distance-expect00}) to calculate the predicted energy consumption of all vehicles after completing the model distribution phase. 
    Based on this observation result, the agent further implements decision-making for selecting the leader in each communication round.
\end{itemize}

For vehicles in the RoI, the corresponding agent is set to active.
If the vehicle has not yet entered the RoI, or once exiting the RoI, the corresponding agent is disabled.

\subsection{Dec-POMDP Reformulation}

In the proposed algorithm, the LSOP problem is reformulated as a Dec-POMDP 
${\langle}\mathcal{I},S,\{\mathcal{A}_i\},T,\mathcal{R},\{\Omega_i\},O,\gamma\rangle$, representing the set of agents, global state set, action set, state transition function, reward function, local observation set, observation probability function, and discount factor, respectively.
In the $k$-th communication round, the observation space, action space, and reward for each agent are as follow.

\begin{enumerate}
    \item \textbf{Observation Space}: The observation space for the local iteration selector of vehicle $i\in\mathcal{I}_r$ consists of the distances and the local round selections of all vehicles:
        \begin{equation}
        \Omega_{\alpha_{ri}} = \left\{ (d_{kiv}, {l}_{k-1,v}) \mid v \in \mathcal{N}, {l}_{kv}\in\mathbb{Z} \right\},i\in\mathcal{I}_r.
        \label{eq:local-obsspace}
        \end{equation}
        For the first communication round, ${l}_{k-1,v}$ is 0. For the leader selector, the observation space $\Omega_{\alpha_l}$ consists of the predicted energy consumption and the remaining energy of all vehicles:
        \begin{equation}
        \Omega_{\alpha_l} = \left\{ ({e'}^{com}_{kv}, e^{res}_{kv}) \mid v \in \mathcal{N} \right\}.
        \label{eq:leader-obsspace}
        \end{equation}
    \item \textbf{Action Space}: For the local iteration selector, its action space $\mathcal{A}_{\alpha_{ri}}$ consists of all possible values for local iteration rounds:
        \begin{equation}
        \mathcal{A}_{\alpha_{ri}} = \left\{ 1, 2, \ldots, \left\lfloor \frac{t^{round}-3T^C_{edge}}{t^{cmp}_{itr}} \right\rfloor \right\},i\in\mathcal{I}_r.
        \label{eq:local-actionspace}
        \end{equation}
        For the leader selector, its action space $\mathcal{A}_{\alpha_{l}}$ consists of the IDs of all vehicles.
        \begin{equation}
        \mathcal{A}_{\alpha_{l}} = \{ \text{ID}_1, \text{ID}_2, \dots, \text{ID}_N \}.
        \label{eq:leader-actionspace}
        \end{equation}
    \item \textbf{Reward}: We set the local iteration selector's reward $\mathcal{R}_{\alpha_{ri}}$ as the number of local iterations:
        \begin{equation}
        \mathcal{R}_{\alpha_{ri}} = {l}_{ki},i\in\mathcal{I}_r.
        \label{eq:local-reward}
        \end{equation}
        For the leader selector, we design a reward function based on the leader selection method in Constraint (\ref{e}). Specifically, the reward $\mathcal{R}_{\alpha_{l}}$ is expressed as
        \begin{equation}
        \mathcal{R}_{\alpha_{l}}=W(\boldsymbol{R}_k)+{\epsilon}G(\boldsymbol{R}_k),
        \label{eq:leader-reward}
        \end{equation}
        where $\epsilon$ represents the reward ratio coefficient.
\end{enumerate}

\subsection{MAPPO-based Joint Optimization of Local Iteration and Leader Selection}

To solve Dec-POMDP, each agent in the proposed algorithm utilizes the MAPPO algorithm for training and decision-making, as detailed in Algorithm \ref{alg:alg1}. 
It is important to note that a single time step within the algorithm corresponds to the execution of one FL communication round in the MDFL environment. 
Upon resetting the MDFL environment, the FL communication round counter is reset to zero, whereas the time step is unaffected.
Each agent trains its policy and value networks using centralized training and decentralized execution to enhance decision quality.
Specifically, each agent maintains the policy network $\pi$ and the value network $V_\phi$.
The policy network outputs a probability distribution over the action space based on the current state, thereby guiding the agent's action.
The value network evaluates the utility of actions by assessing changes in the environment, providing feedback to the policy network to adjust and optimize the action strategy.

\begin{algorithm}[!htbp]
\caption{Training of MAPPO-based joint optimization of local iteration and leader selection}\label{alg:alg1}
\begin{algorithmic}
\STATE 
\STATE {\textbf{Input: }}
$x_v, y_v, s_v, a_v, e^{res}_{kv}, {l}_{k-1,v}, t^{cmp}_{k-1,v}, v\in\mathcal{N}_k$, steps per episode $T_s$, batch size $Z$
\STATE ${l}_{0,v}=0,t^{cmp}_{0,v}=0,v\in\mathcal{N}_k$
\STATE Initialize parameter $\theta_0$ for $\pi_\theta$ and parameter $\phi_0$ for $V_\phi$
\FOR{$k = 1, 2, \dots,$ Episodes}
    \STATE Initialize $D$ as a data buffer for experience replay
    \FOR{$i = 1, 2, \dots, Z$}
        \STATE Initialize $\tau$ as a list for trajectory collection
        \FOR{$t = 1, 2, \dots, T_s$}
            \STATE \# Collect trajectories for $T_s$ communication rounds
            \STATE $a_{i,t}^{(j)} = \pi_{\theta_k}(\Omega_{i,t}^{(j)})$, $j \in \mathcal{I}$
            \STATE $v_{i,t} = V_{\phi_k}(S_{i,t})$
            \STATE Obtain $\boldsymbol{R}_t$ and $\boldsymbol{l}_t$ using $a_{i,t}^{(j)}, j \in \mathcal{I}$
            \STATE Execute a communication round $t$ of MDFL environment using $\boldsymbol{R}_t$ and $\boldsymbol{l}_t$ $\rightarrow r_{i,t}, S_{i,t+1}, \Omega_{i,t+1}$
            \IF{Constraints (\ref{b})-(\ref{f}) are not all satisfied}
                \STATE Reset the MDFL environment
            \ENDIF 
            \STATE $\tau = [\tau; S_{i,t}, \Omega_{i,t}, v_{i,t}, a_{i,t}, r_{i,t}, S_{i,t+1}, \Omega_{i,t+1}]$
            \STATE Compute discount reward $\hat{R}_{i,t}^k$ based on (\ref{discount-reward})
            \STATE Compute advantage $\hat{A}_{i,t}^{k}$ based on (\ref{A-hat})
        \ENDFOR
        \STATE $D = D \cup \tau$
    \ENDFOR
    \STATE Update parameters of $\theta_k$  by (\ref{policy-network}) to obtain $\theta_{k+1}$
    \STATE Update parameters of $\phi_k$ by (\ref{value-network}) to obtain  $\phi_{k+1}$ 
\ENDFOR
\STATE \textbf{return}  $\pi_{\theta}$
\end{algorithmic}
\label{alg1}
\end{algorithm}

During the process of updating the policy network $\theta$, the agent observes the probability $\pi_{\theta}(a_{i,t}^{k}|\Omega_{i,t}^{k})$ of taking action $a$ in local observation $\Omega$ at time $t$ under the current policy $\theta$, and compares it to the action probability $\pi_{\theta_{old}}(a_{i,t}^{k}|\Omega_{i,t}^{k})$ under the previous policy. The ratio of these two probabilities is utilized as the update magnitude for the policy network:

\[
r_{\theta,i,t}^{k}=\frac{\pi_{\theta}(a_{i,t}^{k}|\Omega_{i,t}^{k})}{\pi_{\theta_{old}}(a_{i,t}^{k}|\Omega_{i,t}^{k})}.
\]

The advantage function $A(s, a)$ measures the benefit of taking a specific action in a state relative to the average value of that state. Denote by $\lambda\in[0,1]$ the discount factor, which represents the value discount for future rewards. In PPO, the advantage function is often computed using Generalized Advantage Estimation (GAE):

\begin{align}
    \hat{A}_{i,t}^{k}=\sum_{l=0}^{T_s-t-1}(\gamma\lambda)^{l}\delta_{i,t+l}^{k},
    \label{A-hat}
\end{align}
where $T_s$ represents the timesteps for trajectory collection, and $\delta_{t}^{V}$ represents the temporal difference (TD) error:
\[
\delta_{i,t}^{k}=r_{i,t+l}^k+\gamma V_{\phi}(S_{i,t+1})-V_{\phi}(S_{i,t}),
\]
where $r_{i,t+l}^k$ is the local reward.Additionally, the clipping mechanism constrains the ratio between old and new policies to limit the change in policy updates. An entropy regularization term is included to encourage exploration and avoid premature convergence. The policy objective function is expressed as

\begin{align}
L^{CLIP}(\theta) =& \frac{1}{BnT_s} \sum_{i=1}^{B} \sum_{k=1}^{n} \sum_{t=1}^{T_s}
    \min\left(r_{\theta,i,t}^{k}\hat{A}_{i,t}^{k}, 
     \text{clip}(r_{\theta,i,t}^{k},1-\varepsilon,1+\varepsilon)\hat{A}_{i,t}^{k}\right) \nonumber \\ 
     &+ \sigma \frac{1}{BnT_s} \sum_{i=1}^{B} \sum_{k=1}^{n} \sum_{t=1}^{T_s}
    H\left[\pi_{\theta}(o_{i,t}^{k})\right],
    \label{policy-network}
\end{align}
where $B$ represents the batch size, $n$ represents the number of agents, $H(\cdot)$ represents the entropy of the policy, and $\sigma$ is a hyperparameter that controls the entropy coefficient. The $clip(\cdot)$ function limits the ratio $r_t$ to the range $[1-\varepsilon,1+\varepsilon]$, ensuring more stable updates. When the advantage function is positive, it means the new policy is better, and $r_t$ should be larger. However, if $r_t>1+\varepsilon$, no extra benefit is provided. If the advantage function is negative, a smaller $r_t$ favors the old policy, but no additional retention mechanism applies when $r_t<1-\varepsilon$. This approach keeps the difference between the old and new policies within a reasonable range.

The value network $V_{\phi}$ learns the mapping from state to reward, aiming to predict the expected return for each state by estimating the value function $V(\cdot)$.
The objective function of $\phi$ is expressed as

\begin{align}
L^{CLIP}(\phi) = \frac{1}{BnT_s} \sum_{i=1}^B \sum_{k=1}^n \sum_{t=1}^{T_s}
\max \left[ \left( V_\phi ( S_{i,t} ) - \hat{R}_{i,t}^k \right)^2, \right. \nonumber \\
\left. \left( {clip} ( {V}_{\phi} ( S_{i,t} ), 
{V}_{\Phi_{\text{old}}} ( S_{i,t} ) - \varepsilon, 
{V}_{\Phi_{\text{old}}} ( S_{i,t} ) + \varepsilon ) 
- \hat{R}_{i,t}^k \right)^{2} \right].
\label{value-network}
\end{align}
$\hat{R}_i^k$ is the discount reward,
\begin{align}
\hat{R}_{i,t}^k=\sum_{l=0}^{T_s-t-1}(\gamma^lr_{i,t+l}^k),
\label{discount-reward}
\end{align}

\subsection{Complexity Analysis}

The complexity of deep reinforcement learning (DRL) mainly comes from the forward propagation during action prediction and the backward propagation during network training \cite{ning2024multiple}.
Assume that the actor network $\pi_{\theta_{\alpha}}, \forall \alpha \in \mathcal{I}$, has $Q$ fully connected layers.
The $q$-th layer has $\omega^q_{\theta_{\alpha}}$ neural nodes, $q \in \{1,2,\dots,Q\}$.
In the actor network $\pi_{\theta_{\alpha}}$, $\forall \alpha \in \mathcal{I}$, denote by $\omega^r_{\theta_{\alpha}}$ and $\omega^t_{\theta_{\alpha}}$ the sum of nodes in the ReLU layers and the tanh layers, respectively.
The results of $\omega^q_{\phi_{\alpha}}$ and $\omega^r_{\phi_{\alpha}}$ in the value network $V_{\phi}$ are consistent with that of the actor network.
In the action prediction process, for agent $\alpha$, the complexity caused by the forward propagation is $\mathcal{O}(\sum_{q=1}^Q \omega^q_{\theta_\alpha} \omega^{q+1}_{\theta_\alpha} + \sum_{q=1}^Q \omega^q_{\phi_\alpha} \omega^{q+1}_{\phi_\alpha})$ \cite{zhong2022hybrid}.
During the agent training process, the complexities caused by both forward and back propagation are $\mathcal{O}(\sum_{q=1}^Q \omega^q_{\theta_\alpha} \omega^{q+1}_{\theta_\alpha} + \sum_{q=1}^Q \omega^q_{\phi_\alpha} \omega^{q+1}_{\phi_\alpha})$ and $\mathcal{O}(\omega_{\theta_\alpha}^r+6\omega_{\theta_\alpha}^t+\omega_{\phi_\alpha}^r+6\omega_{\phi_\alpha}^t+\sum_{q=1}^Q\omega_{\theta_\alpha}^q\omega_{\theta_\alpha}^{q+1}+\sum_{q=1}^Q\omega_{\phi_\alpha}^q\omega_{\phi_\alpha}^{q+1})$ \cite{zhong2022hybrid}.
We normally set $\omega^q_{\theta_\alpha} \gg 6, \forall \alpha \in \mathcal{I}, \forall q$ \cite{ning2024multiple}.
For the GAE, the complexity is $\mathcal{O}(I^2\!L)$ \cite{kang2023cooperative}, where $I$ represents the number of agents, and $L$ represents the maximum length of the feature vector observed by the agent.
Thus, the overall computational complexity of the proposed algorithm in an episode can be expressed as $\mathcal{O}(\sum_{\alpha\in\mathcal{I}}2T_s(Z+1)(\sum_{q=1}^Q\omega_{\theta_\alpha}^q\omega_{\theta_\alpha}^{q+1}+\sum_{q=1}^Q\omega_{\phi_\alpha}^q\omega_{\phi_\alpha}^{Q+1}))+\mathcal{O}( I^2\!LT_s)$, where $Z$ represents the batch size.

\section{Experiment}

\subsection{Experiment Setting}

\begin{enumerate}
    \item \textbf{Dataset and learning task}: 
    We utilize the LeNet network for training on the Fashion MNIST dataset.
    Fashion MNIST consists of 10 categories, each corresponding to a different fashion item with 6,000 training images and 1,000 test images\cite{xiao2017}.
   LeNet, a classical convolutional neural network, exemplifies the effectiveness of CNNs in image classification, as evidenced by its outstanding performance in handwritten digit recognition tasks \cite{lenet}.
    \item \textbf{Parameter setup}: The RoI is set as a rectangular area with a length of 3 km in the east-west direction and 200 m in the north-south direction. The main road goes from west to east, and this direction is designated as the x-axis. The main road has three lanes, and the ramp has one lane. The origin is located at the westernmost side of the main road. There are three exit ramps at 500 m, 1 km, and 2.5 km from the origin, and an entrance ramp at 2 km. Vehicles enter the RoI from the west side of the main road and the entrance ramp, and exit the RoI from the east side of the main road and the exit ramps, resulting in six different driving routes. The routes and the specific entry times into the RoI are randomly sampled using a uniform distribution, and traffic flow and vehicle data within the RoI are generated using the traffic simulation software SUMO. 
    The work in \cite{Pari2023} compares the packet delivery rate (PDR) of vehicle-to-vehicle communication using DSRC and LTE-V2X technologies. 
    The reliable communication range of LTE-V2X is between 205 meters and 1,000 meters, achieving at least 90\% PDR, while DSRC has a reliable communication range of 270 meters to 1,025 meters, achieving the same PDR. 
    Therefore, in this paper, the radius $r$ of the vehicle direct communication area is set to 200 m. The initial energies of vehicles $E_v$, $e^{cmp}_{itr}$, $E^{C}_{edge}$, and $E^{C}_{cloud}$ are set to 1,000 units, 5 units, 2 units, and 5 units, respectively. 
    The time duration of $t^{round}$, $t^{cmp}_{itr}$, $T^{C}_{edge}$, and $T^{C}_{cloud}$ are set to 10 units, 1 unit, 1 unit, and 2 units, respectively. 
    The reward ratio coefficient is set to 1.
    The number of vehicles is set to 10.
    Parameter $\rho$ is set to 5.
    The number of episodes is set to 10,000. 
    Parameter $T_s$ is set to 100. The clipping parameter $\varepsilon$ is set to 0.2.
    The discount factor $\lambda$ is set to 0.99. The learning rate for both the actor and critic is set to 0.0005.
    In our experiments, the evaluation criteria include accuracy $F^{acc}$ and energy consumption ratio, denoted by $ECR$. The average accuracy $F^{acc}$ is defined in (\ref{eq:final-accuracy}).
    $ECR$ is defined as
     \begin{align}
    ECR=\sum_{v\in\mathcal{N}}\frac{\sum_{k=1}^{K}e^{cmp}_{kv}}{\sum_{k=1}^{K}e^{sum}_{kv}+\zeta},
    \end{align}
    where $\zeta$ is a small positive number to prevent division by zero.
    Denote by $e^{total}_{k}$ the total energy consumption of the MDFL framework in the $k$-th communication round, which is expressed as
    \begin{align}
    e^{total}_{k}=\sum_{v\in\mathcal{N}}(E_v-e^{res}_{kv}).
    \end{align}

    \item \textbf{Baseline}: We compare the proposed algorithm with the following two algorithms.
        \begin{itemize}
            \item \textbf{Random}: The leader is randomly selected, and vehicles independently randomly select the local round.
            \item \textbf{DFL}: Referencing the DFL framework proposed in \cite{Hu2019DFL}, the algorithm treats all vehicles in the RoI as neighbors and does not select the leader. The local iteration is fixed as $\lfloor\frac{t^{round}-2T^C_{edge}}{t^{cmp}_{itr}} \rfloor$.
        \end{itemize}
\end{enumerate}


\subsection{Comparison With Respect To Various Settings}

\begin{figure}[!htbp]
    \centering
    \begin{tikzpicture}
    \begin{axis}[  
        xlabel=$E_v$,  
        ylabel=$F^{acc}$,  
        enlarge x limits=false, 
        legend style={at={(0.95,0.35)},anchor=south east}, 
        legend cell align={left}, 
        line width=1.2pt, 
    ]  
      
    \addplot table[x=x, y=mappo, col sep=comma] {plot/energy-acc.csv};  
    \addlegendentry{MAPPO} %
      
    \addplot table[x=x, y=dfl, col sep=comma] {plot/energy-acc.csv};  
    \addlegendentry{DFL} 
    
    \addplot table[x=x, y=random, col sep=comma] {plot/energy-acc.csv};  
    \addlegendentry{Random}
      
    \end{axis}  
    \end{tikzpicture}
    \caption{Comparison of $F^{acc}$ with respect to initial vehicle energy level.}
    \label{fig:energy-acc}
\end{figure}
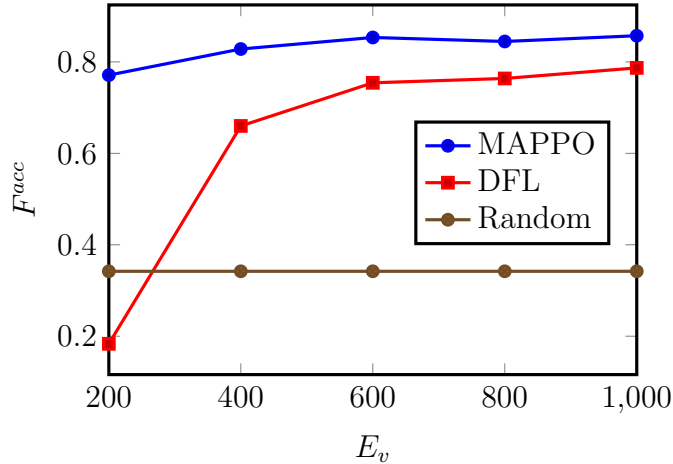

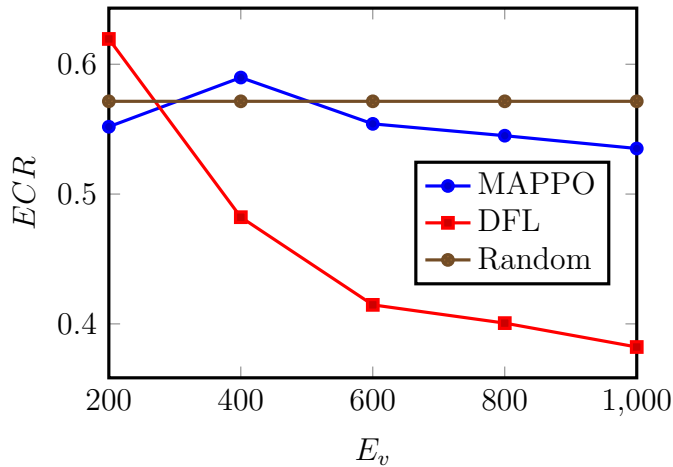
\begin{figure}[!htbp]
    \centering
    \begin{tikzpicture}
    \begin{axis}[  
        xlabel=$E_v$,  
        ylabel=$ECR$,  
        enlarge x limits=false, 
        legend style={at={(0.95,0.25)},anchor=south east}, 
        legend cell align={left}, 
        line width=1.2pt, 
    ]  
      
    \addplot table[x=x, y=mappo, col sep=comma] {plot/energy-rate.csv};  
    \addlegendentry{MAPPO} %
      
    \addplot table[x=x, y=dfl, col sep=comma] {plot/energy-rate.csv};  
    \addlegendentry{DFL} 
    
    \addplot table[x=x, y=random, col sep=comma] {plot/energy-rate.csv};
    \addlegendentry{Random}
      
    \end{axis}  
    \end{tikzpicture}
    \caption{Comparison of $ECR$ with respect to initial vehicle energy level.}
    \label{fig:energy-rate}
\end{figure}

Figs. \ref{fig:energy-acc} and \ref{fig:energy-rate} show the impact of the initial energy $E_v$ on the proposed algorithm. 
We can gain the following insights. 
First, the proposed algorithm is better than the other algorithms. 
This is because the proposed algorithm effectively selects local iterations and leaders, 
enabling increased communication rounds and subsequently maximizing resource utilization.
Second, with the increase of $E_v$, the accuracy gap between DFL and the proposed algorithm decreases, but the $ECR$ of DFL decreases significantly.
When $E_v=200$, the $F^{acc}$ of the proposed algorithm is 125\% higher than Random and 320\% higher than DFL with a comparable $ECR$. When $E_v=1000$, the $F^{acc}$ of the proposed algorithm is 9\% higher than DFL, and the $ECR$ is 40\% higher than DFL.
The reason is that with the increase of $E_v$, the constraint on energy reduces. 
In DFL, each vehicle needs to exchange model parameters with other vehicles in the RoI, leading to rapid consumption of energy resources.
As more vehicles are added, the energy consumption of each communication round increases rapidly, resulting in a gradual decline in the $ECR$, and eventually an early exit due to the exhaustion of energy. 
In addition, when $E_v=400$, the proposed algorithm heuristically discovered a more energy-efficient selection scheme, achieving a balance between energy ratio and participation ratio, which resulted in an improved $ECR$.

Finally, the $F^{acc}$ and $ECR$ of the random algorithm remain constant with the increase of $E_v$. 
This is because it inadequately accounts for constraints when determining local iteration rounds and selecting leaders.
Consequently, it frequently terminates prematurely due to constraint violations. 
The premature termination of this algorithm is not directly correlated with the initial energy level of the vehicle.

\begin{figure}[!htbp]
    \centering
    \begin{tikzpicture}
    \begin{axis}[  
        xlabel=$E^C_{cloud}$,  
        ylabel=$ECR$,  
        enlarge x limits=false, 
        legend style={at={(0.95,0.55)},anchor=south east}, 
        legend cell align={left}, 
        line width=1.2pt, 
    ]  
      
    \addplot table[x=x, y=mappo, col sep=comma] {plot/cloud-energy-rate.csv};  
    \addlegendentry{MAPPO} %
      
    \addplot table[x=x, y=dfl, col sep=comma] {plot/cloud-energy-rate.csv};  
    \addlegendentry{DFL} 
    
    \addplot table[x=x, y=random, col sep=comma] {plot/cloud-energy-rate.csv};
    \addlegendentry{Random}
      
    \end{axis}  
    \end{tikzpicture}
    \caption{Comparison of $ECR$ with respect to indirect transmission energy consumption.}
    \label{fig:cloud-energy-rate}
\end{figure}
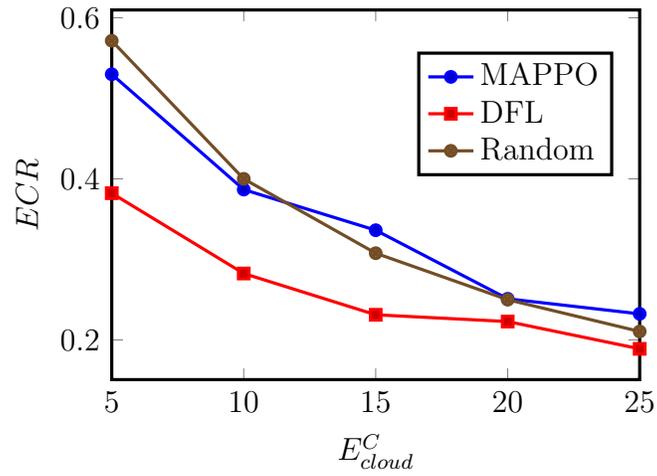

Figs. \ref{fig:cloud-energy-acc} and \ref{fig:cloud-energy-rate} show the impact of the indirect transmission energy $E^C_{cloud}$ on the proposed algorithm.
We gain the following insights.  
First, the proposed algorithm outperforms DFL.
When $E^C_{cloud}=25$, the $F^{acc}$ of the proposed algorithm is 26\% higher than DFL, while the energy consumption rate is 22\% higher.
This is because the proposed algorithm remains able to select the optimal leader, even when the indirectly transmitted energy increases, ensuring sufficient communication rounds and a high $ECR$.
Second, with the increase of $E^C_{cloud}$, 
the proposed algorithm remains relatively stable, while DFL decreases in $F^{acc}$.
This is because of the higher energy consumption cost associated with increased $E^C_{cloud}$, and the vehicle has a higher probability of early exit due to the exhaustion of energy. 
DFL is more sensitive to energy consumption. 
This phenomenon is amplified by the increase of $E^C_{cloud}$. 
Finally, with the increase of $E^C_{cloud}$, the gap between the $ECR$ of the proposed algorithm and that of DFL shrinks, and the $ECR$ of both algorithms decreases as $E^C_{cloud}$ increases.
Note that the proposed algorithm accounts for vehicle participation ratios, which result in some energy loss that becomes more pronounced as $E^C_{cloud}$ increases.

\begin{figure}[!htbp]
    \centering
    \begin{tikzpicture}
    \begin{axis}[  
        xlabel=$E^C_{cloud}$,  
        ylabel=$F^{acc}$,  
        enlarge x limits=false, 
        legend style={at={(0.95,0.15)},anchor=south east}, 
        legend cell align={left}, 
        line width=1.2pt, 
    ]  
      
    \addplot table[x=x, y=mappo, col sep=comma] {plot/cloud-energy-acc.csv};  
    \addlegendentry{MAPPO} %
      
    \addplot table[x=x, y=dfl, col sep=comma] {plot/cloud-energy-acc.csv};  
    \addlegendentry{DFL} 
    
    \addplot table[x=x, y=random, col sep=comma] {plot/cloud-energy-acc.csv};  
    \addlegendentry{Random}
      
    \end{axis}  
    \end{tikzpicture}
    \caption{Comparison of $F^{acc}$ with respect to indirect transmission energy consumption.}
    \label{fig:cloud-energy-acc}
\end{figure}
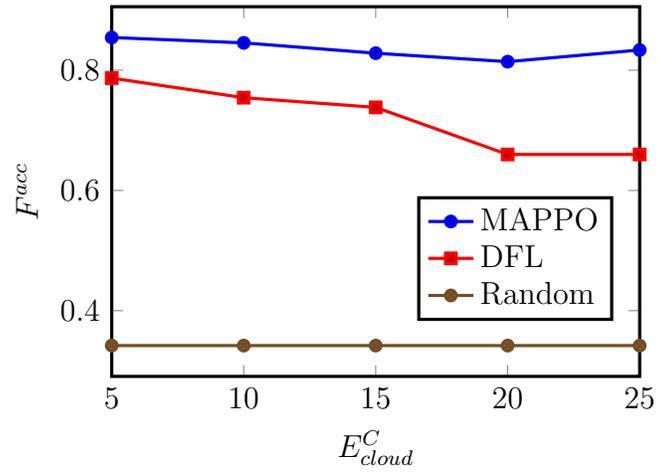

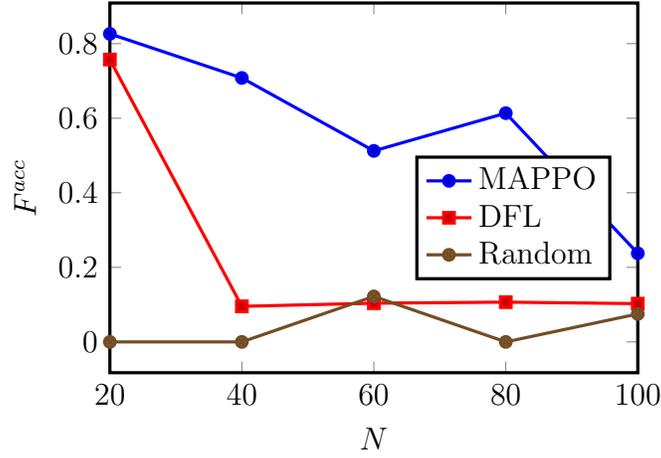
\begin{figure}[!htbp]
    \centering
    \begin{tikzpicture}
    \begin{axis}[  
        xlabel=$N$,  
        ylabel=$F^{acc}$,  
        enlarge x limits=false, 
        legend style={at={(0.95,0.25)},anchor=south east}, 
        legend cell align={left}, 
        line width=1.2pt, 
    ]  
      
    \addplot table[x=x, y=mappo, col sep=comma] {plot/vehicle-acc.csv};  
    \addlegendentry{MAPPO} %
      
    \addplot table[x=x, y=dfl, col sep=comma] {plot/vehicle-acc.csv};  
    \addlegendentry{DFL} 
    
    \addplot table[x=x, y=random, col sep=comma] {plot/vehicle-acc.csv};  
    \addlegendentry{Random}
      
    \end{axis}  
    \end{tikzpicture}
    \caption{Comparison of $F^{acc}$ with respect to the number of vehicles.}
    \label{fig:vehicle-acc}
\end{figure}

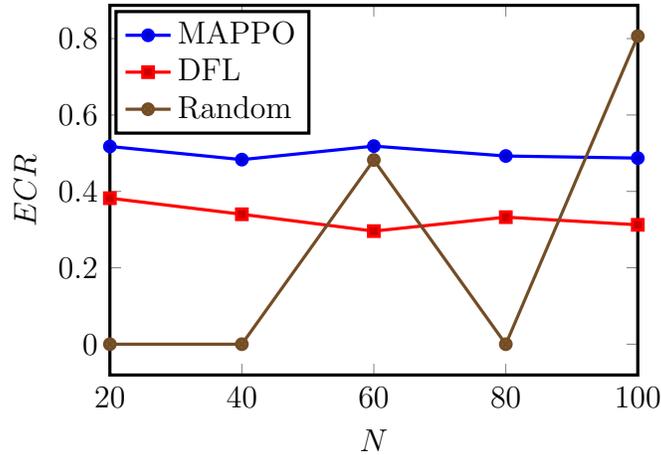
\begin{figure}[!htbp]
    \centering
\begin{tikzpicture}
    \begin{axis}[  
        xlabel=$N$,  
        ylabel=$ECR$,  
        enlarge x limits=false, 
        legend style={at={(0.38,0.65)},anchor=south east}, 
        legend cell align={left}, 
        line width=1.2pt, 
    ]  
      
    \addplot table[x=x, y=mappo, col sep=comma] {plot/vehicle-rate.csv};  
    \addlegendentry{MAPPO} %
      
    \addplot table[x=x, y=dfl, col sep=comma] {plot/vehicle-rate.csv};  
    \addlegendentry{DFL} 
    
    \addplot table[x=x, y=random, col sep=comma] {plot/vehicle-rate.csv};
    \addlegendentry{Random}
      
    \end{axis}  
    \end{tikzpicture}
    \caption{Comparison of $ECR$ with respect to the number of vehicles.}
    \label{fig:vehicle-rate}
\end{figure}

Figs. \ref{fig:vehicle-acc} and \ref{fig:vehicle-rate} illustrate the impact of the number of vehicles on the proposed algorithm.
We gain the following insights.
First, MAPPO consistently outperforms the other algorithms, achieving at least 9\% higher for $F^{acc}$ and 34\% higher for $ECR$ than DFL.
However, the efficiency of MAPPO exhibits fluctuations and a downward trend.
As the number of vehicles increases, the algorithm's solution space becomes more complex, leading to decreased stability in its efficiency. Additionally, with more vehicles, the amount of data in each vehicle's local dataset diminishes.
Since vehicles enter the RoI sequentially,  the algorithm utilizes less data, resulting in a decline in MAPPO's $F^{acc}$.
Second, DFL fails when $N \geq 40$.
This is because the newly added vehicles do not have enough data for training, resulting in underfitting and overwriting of model weight updates.
The leader-based model distribution mechanism helps mitigate this issue, allowing MAPPO to be less affected. Lastly, the Random algorithm is unable to find a valid solution when $N \in \{ 20,40,80 \}$.
As $N$ increases, the solution space of Eq. (\ref{eq:final-target}) becomes more intricate, reducing the likelihood of extracting a valid solution by the Random algorithm.


\subsection{Convergence Performance of MAPPO}

\begin{figure}[ht]
\centering
\includegraphics[width=0.85\textwidth]{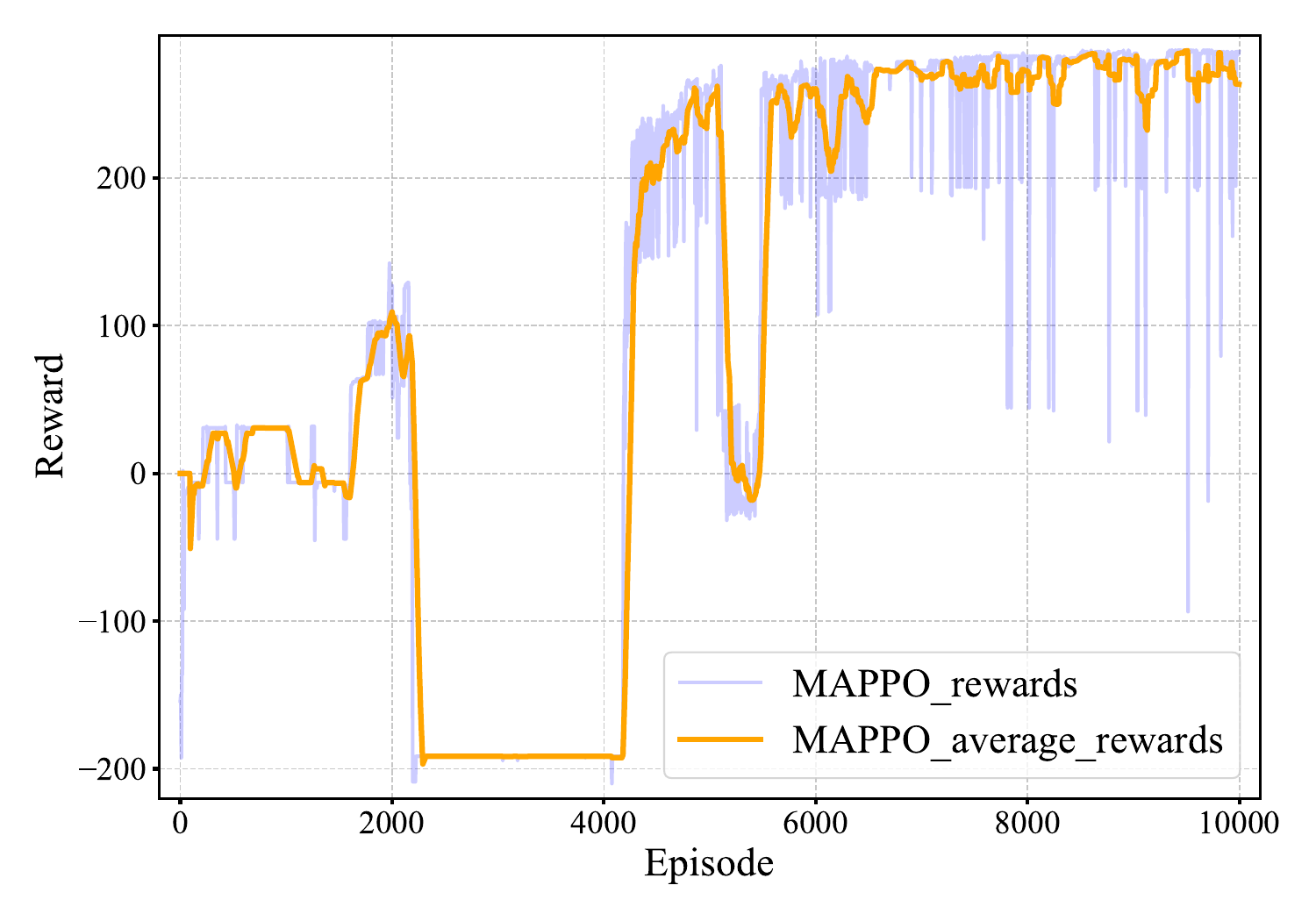}
\begin{center}
\caption{Convergence performance of the accumulated rewards.}
\label{fig:convergence}
\end{center}
\end{figure}

LSOP is a complex combinatorial optimization problem, and the optimization variables $\boldsymbol{R}$ and $\boldsymbol{l}$ are coupled.
Directly solving the proposed optimization problem is challenging \cite{feng2023qoe}.
Some works have proposed MAPPO-based algorithms to solve complex combinatorial optimization problems \cite{kuba2021trust,nie2024asynchronous}.
They have demonstrated that MAPPO exhibits certain convergence properties.
These convergence properties guarantee the stability and performance of MAPPO when dealing with complex combinatorial optimization problems.
Drawing on the analysis in \cite{ning2024multiple}, we conduct an analysis of the convergence performance of the proposed algorithm, as depicted in Fig. \ref{fig:convergence}.
It can be observed that the algorithm converges in approximately $7\times10^3$ episodes.
Moreover, the proposed algorithm becomes trapped in invalid solutions at $2\times10^3$ and $5\times10^3$ episodes but subsequently escapes from these solutions. This behavior can be attributed to the exploratory nature of the entropy regularization term.
Finally, after $7\times10^3$ episodes, the reward value of the proposed algorithm oscillates. This is a result of the exploration introduced by the entropy regularization term. Meanwhile, the clipping strategy ensures that the policy does not change too rapidly, enabling the proposed algorithm to maintain stability.

\subsection{Sensitivity Analysis}

\begin{figure}[!htbp]
    \centering
    \begin{tikzpicture}
    \begin{axis}[  
        xlabel=$e^{total}_k$,  
        ylabel=Accuracy of each round,
        enlarge x limits=false, 
        legend style={at={(0.98,0.05)},anchor=south east}, 
        legend cell align={left}, 
        line width=1.2pt, 
    ]  
    
    \addplot+[color=brown] table[x=x1, y=y1, col sep=comma] {plot/reward-acc.csv};  
    \addlegendentry{$\epsilon=0$} %

    \addplot+[color=red] table[x=x2, y=y2, col sep=comma] {plot/reward-acc.csv};  
    \addlegendentry{$\epsilon=0.01$} %

    \addplot+[color=blue] table[x=x3, y=y3, col sep=comma] {plot/reward-acc.csv};  
    \addlegendentry{$\epsilon=0.1$} %

    \addplot+[color=purple] table[x=x4, y=y4, col sep=comma] {plot/reward-acc.csv};  
    \addlegendentry{$\epsilon=1$} %

    \addplot+[color=orange] table[x=x5, y=y5, col sep=comma] {plot/reward-acc.csv};  
    \addlegendentry{$\epsilon=10$} %

    \end{axis}  
    \end{tikzpicture}
    \caption{Accuracy curves with respect to reward ratio coefficient.
    }
    \label{fig:reward-acc}
\end{figure}
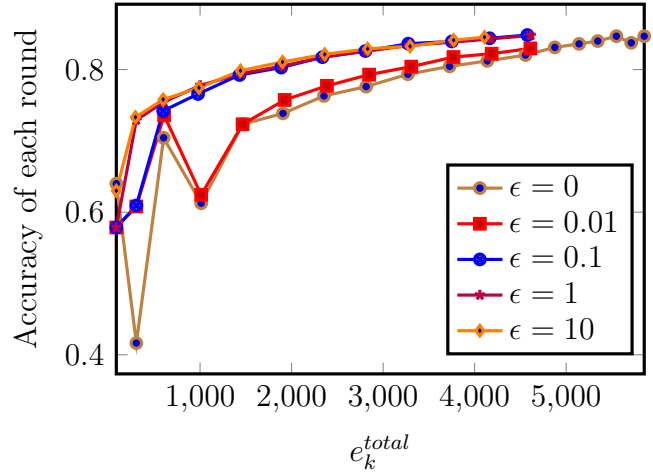

Fig. \ref{fig:reward-acc} and Tab. \ref{tab:ecr-reward-rate} demonstrate the impact of the reward proportional coefficient $\epsilon$ on the performance.
We can see that with the increase of $\epsilon$, the accuracy of the proposed algorithm becomes smoother.
When $\epsilon = 0.01$, the accuracy initially drops by about 15\%, then increases by 16\%, almost returning to its original level. When $\epsilon \geq 0.1$, this phenomenon disappears.
This is because $\epsilon$ represents the impact of participation ratio on the agent's reward. 
A higher $\epsilon$ means that the agent is more inclined to choose the vehicles that have entered the RoI in recent communication rounds as the leader, rather than those participating in training for the first time.
In addition, with the increase of $\epsilon$, the communication rounds required by the proposed algorithm decreases.
This is because $\epsilon$ reflects the balance between the ratio of energy proportion and that of participation.
When $\epsilon$ is small, the selected leader is not the optimal one, but it achieves a more balanced energy consumption among vehicles, and enables a larger number of communication rounds. 
This setting is suitable for situations where the number of network nodes $N$ fluctuates to a relatively small extent. 
With a high $\epsilon$ setting, the proposed algorithm effectively identifies and selects suitable leaders while significantly reducing the need for communication rounds. 
This feature is especially applicable to scenarios where the number of network nodes $N$ fluctuates greatly, allowing it to effectively cope with dynamic changes in the environment. 
Finally, it is important to highlight that no significant fluctuations in the system $ECR$ were observed across the tested $\epsilon$ values.
This reveals the independence in function execution between the local iteration selector and the leader selector. 
Modifications to the leader selector's internal reward formula has a negligible impact on the operation of the local iteration selector.

Tab. \ref{tab:communication_range} presents the imapct of communication range on MAPPO. 
It can be observed that the $F^{acc}$ of MAPPO remains relatively stable, 
indicating that MAPPO exhibits robust convergence across different direct communication ranges.
In addition, as the direct communication range increases, the $ECR$ of MAPPO also increases, indicating that MAPPO is capable of utilizing energy in accordance with the direct communication range.

\begin{table}[htbp]
    \caption{Energy consumption ratio $ECR$ with respect to reward ratio coefficient.}
    \centering
    \begin{tabular}{|l||c|}
        \hline
        $\epsilon$ & $ECR$ \\ \hline
        0          & 0.5459 \\ \hline
        0.01       & 0.5457 \\ \hline
        0.1        & 0.5332 \\ \hline
        1          & 0.5343 \\ \hline
        10         & 0.5355 \\ \hline
    \end{tabular}
    \label{tab:ecr-reward-rate}
\end{table}

\begin{table}[htbp]
    \caption{The performance of MAPPO with respect to direct communication range.}
    \centering
    \begin{tabular}{|l||c|c|}
        \hline
        $r$ (M) & $F^{acc}$ & $ECR$ \\ \hline
        100 & 0.8602 & 0.5037 \\ \hline
        200 & 0.8667 & 0.5115 \\ \hline
        300 & 0.8820 & 0.5785 \\ \hline
        400 & 0.8654 & 0.6293 \\ \hline
        500 & 0.8795 & 0.6166 \\ \hline
    \end{tabular}
    \label{tab:communication_range}
\end{table}

\subsection{Impact of Non-IID Data on MDFL}

\begin{figure}[ht]
    \centering
    \includegraphics[width=1\textwidth]{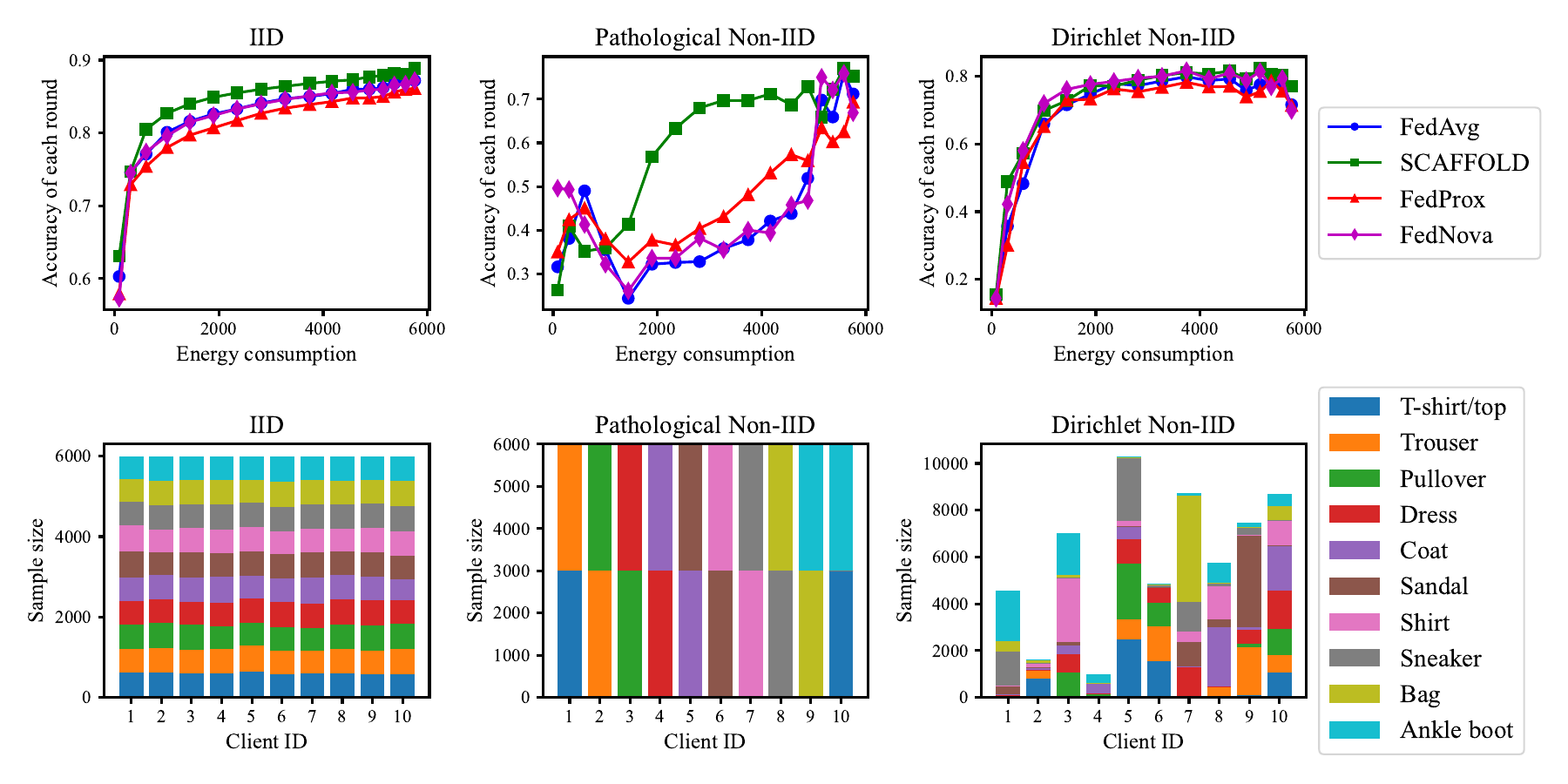}
    \begin{center}
    \caption{Accuracy curves of MAPPO with respect to aggregation strategies, along with the corresponding data distributions under IID, Pathological non-IID, and Dirichlet non-IID.}
    \label{fig:distribution}
    \end{center}
\end{figure}

We divide the FashionMNIST dataset in the following two different ways, namely Pathological non-IID (class imbalance distribution) and Dirichlet non-IID (data imbalance distribution).
In the Pathological non-IID data, the data for each class is divided into two equal parts, each vehicle being assigned one part from the two classes \cite{McMahan2017}.
In the Dirichlet non-IID data, the data of each class is divided into 10 clients according to a Dirichlet distribution, with the concentration parameter set to 0.3 \cite{kim2022multi}.
We evaluate four aggregation strategies, i.e., FedAvg, FedNova, FedProx, and Scaffold.
The experimental results are presented in Fig. \ref{fig:distribution}.
FedAvg is a classical algorithm in FL with the advantages of simple implementation, and easy deployment\cite{McMahan2017}. 
To address issues caused by system heterogeneity, FedNova effectively resolves the aggregation bias that may occur with FedAvg when local gradient descent iterations are inconsistent, by introducing a normalized update method, thereby enhancing the algorithm's adaptability and stability\cite{Wang2020FedNova}.
FedProx adds the proximal term to the local optimization process to alleviate the problem of large update differences between clients caused by data heterogeneity\cite{Li2020FedProx}.
Scaffold uses control variables to compensate for update differences between clients. 
This method reduces the negative impact of Non-IID data on FL. 
By correcting the update direction of clients, Scaffold also improves convergence performance while reducing the number of communication rounds\cite{Karimireddy2020SCAFFOLD}. 

The heterogeneity of the MDFL framework is mainly reflected in two aspects. 
First, due to the energy and time constraints, the local iteration selection of vehicles is not exactly the same. 
In addition, due to changes in membership of $\mathcal{N}_k$, the data set in the RoI also changes. The two aspects result in system and data heterogeneity.
In the IID data scenario, all evaluated aggregation strategies display a smooth rising accuracy curve, indicating that the impact of system heterogeneity on the performance of these strategies is limited. In particular, Scaffold achieves the highest $F^{acc}$ in this scenario and is least affected by system heterogeneity. 
However, in the Non-IID data scenario, the operation process of the algorithm shows significant fluctuations in accuracy, and $F^{acc}$ decreases compared with the IID scenario.
It can be observed that $F^{acc}$ of the four aggregation strategies decreases by an average of 19\% under the Pathological Non-IID scenario and 17\% under the Dirichlet Non-IID scenario compared to the IID scenario.
This is because in the Dirichlet Non-IID scenario, the imbalance in data distribution leads to inconsistent directions of local update gradients, resulting in a decrease in $F^{acc}$.
In the Pathological Non-IID scenario, the data corresponding to different labels are not evenly distributed, resulting in some late-appearing label data that are not visible to the model. 
As the training process progresses, the model gradually adapts and correctly identifies these late-appearing data.

Among the four aggregation strategies, Scaffold demonstrates superior performance in both IID and Non-IID scenarios.
 The control variables in Scaffold encompass the gradients from all clients. 
 When a client conducts local updates, it accounts for the gradient adjustments of other clients, thereby ensuring that the gradients for update remain consistent across all clients. 
 Consequently, Scaffold approximates the ideal condition of employing unbiased gradient updates via these control variables. 
 This characteristic empowers Scaffold to effectively counteract client drift induced by heterogeneity.
FedAvg performs a weighted average based on the number of samples held by each client. It lacks sufficient ability to handle heterogeneity.
FedNova mitigates system heterogeneity caused by inconsistent local iteration counts by using a normalized averaging method.
However, this approach is affected by data heterogeneity.
The regularization term in FedProx penalizes clients that deviate significantly from the global model.
In MDFL, the leader for each communication round is dynamically selected, which results in the inferior performance of this regularization term.

\subsection{Generalization Ability of MDFL}

\begin{figure}[ht]
\centering
\includegraphics[width=0.9\textwidth]{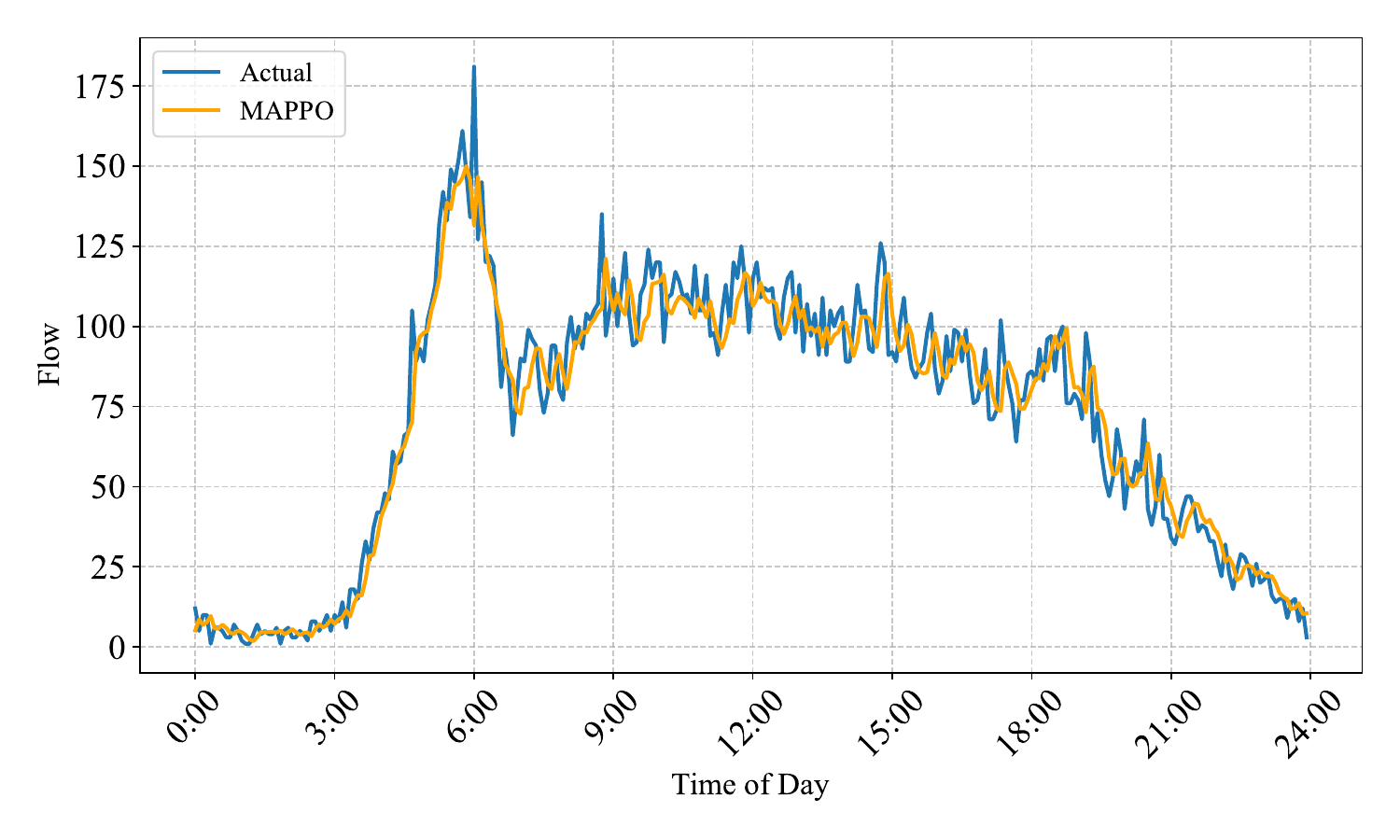}
\begin{center}
\caption{Traffic flow prediction of MAPPO.}
\label{fig:pems}
\end{center}
\end{figure}

To evaluate the generalization ability of MDFL, we apply MDFL to the traffic flow prediction (TFP) task.
This task entails training a gated recurrent unit (GRU) model using a time-series data from the Caltrans Performance Measurement System (PeMS) dataset \cite{chen2002freeway}.
The experimental setup follows the work in \cite{liu2020privacy}.
The PeMS dataset compiles traffic flow data from over 39,000 individual detectors across the freeway systems in California's metropolitan areas \cite{lv2014traffic}.
For our experiment, we use traffic flow data gathered from January to March 2013.
Specifically, the traffic data from January to February serves as the training dataset, while the data from March serves as the test dataset.
We use mean absolute error (MAE), root mean square error (RMSE), and mean absolute percentage error (MAPE) to evaluate the prediction accuracy, which are defined as
\begin{align}
    \left\{
    \begin{aligned}
        \mathrm{MAE} &= \frac{1}{M} \sum\limits_{m=1}^{M} {|\mu_{m} - \hat{\mu}_{m}|}, \\
        \mathrm{RMSE} &= \left[ \frac{1}{M} \sum\limits_{m=1}^{M} {\left( |\mu_{m} - \hat{\mu}_{m}| \right)^{2}} \right]^{\frac{1}{2}}, \\
        \mathrm{MAPE} &= \frac{100\%}{M} \sum\limits_{m=1}^{M} \left| \frac{\hat{\mu}_{m} - \mu_{m}}{\mu_{m}} \right|,
    \end{aligned}
    \right.
    \nonumber
\end{align}
where $M$ represents the sample size, $\mu$ represents the actual traffic flow, and $\hat{\mu}$ represents the predicted traffic flow.
In TFP, MAE quantifies the absolute difference between the predicted and actual traffic flow values.
RMSE emphasizes the model's performance during periods of high traffic or when there are significant deviations between predictions and actual values, and MAPE measures the average magnitude of the relative differences between the predicted and actual traffic flow values.
The results of the metric analysis are presented in Table \ref{tab:pems}.
In Table \ref{tab:pems}, the CGRUL algorithm refers to training the GRU model for 100 iterations on the entire training set without splitting the dataset.
The experimental results of MAPPO are competitive with respect to CGRUL, which indicates that MDFL achieves comparable training performance to centralized training in TFP.
In addition, Fig. \ref{fig:pems} illustrates the predicted traffic flow for one day on the test dataset.
The prediction results of MAPPO accurately reflect the actual traffic flow characteristics for the entire 24-hour period, which indicates that the GRU model trained with MAPPO can effectively predict traffic flow.

\begin{table}[ht]
    \caption{Performance Comparison Between the MDFL Framework and Centralized GRU Learning (CGRUL).}
    \centering
    \begin{tabular}{|l||c|c|c|}
        \hline
        Algorithm & MAE & RMSE & MAPE \\ \hline
        MAPPO & 7.33 & 10.00 & 18.33\% \\ \hline
        CGRUL & 7.20 & 9.80 & 18.34\% \\ \hline
    \end{tabular}
    \label{tab:pems}
\end{table}

\section{Conclusion}

In this paper, we propose a mobility-aware decentralized federated learning framework for vehicular networks. 
Due to vehicle mobility, the vehicles participating in the training of the federated learning model undergo dynamic changes.
We model a local iteration and leader selection joint optimization problem, i.e., LSOP, considering the impact of vehicle mobility and resource constraints. 
To solve this problem, we first reformulate LSOP as a Dec-POMDP problem, 
and then propose an effective algorithm based on MAPPO to solve Dec-POMDP. 
We conduct comprehensive experiments to validate the impact of various parameters, including initial vehicle energy, indirect transmission energy consumption, and the number of vehicles, on the proposed algorithm. 
The experiment results manifest that the proposed algorithm outperforms the existing algorithms, and achieves a reasonable trade-off between energy consumption and participation ratio. 
Additionally, we evaluate the impact of aggregation strategies on the proposed algorithm. 
The results indicate that, for the same setup, the Scaffold strategy handles system heterogeneity in a more effective and efficient manner. 


\bibliographystyle{elsarticle-num} 
\bibliography{ref}

\end{document}